\documentclass[rnote]{aa}
\usepackage{psfig,latexsym,graphicx,natbib} 
\bibpunct{(}{)}{;}{a}{}{,}
\pssilent
\begin{document}
\title{The most rapidly rotating He-strong emission line star: HR\,7355\thanks{Based on
    observations collected at La Silla, ESO-Chile
    (Prop.\ 063.H-0080, 073.D-0234)}}
\author{Th.~Rivinius\inst{1}
\and S.~\v{S}tefl\inst{1}
\and R.H.D.~Townsend\inst{2}
\and D.~Baade\inst{3}
}
\offprints{Th.\ Rivinius\\
\email{tiviniu@eso.org}}
\institute{ESO - European Organisation for Astronomical Research in the Southern Hemisphere, Chile
\and Bartol Research Institute, University of Delaware, Newark, DE 19716, USA
\and ESO - European Organisation for Astronomical Research in the Southern Hemisphere, Karl-Schwarzschild-Str.~2, 85748 
Garching bei M\"unchen, Germany 
}
\date{Received: $<$date$>$; accepted: 24/01/2008}
\abstract
{}
{We searched for massive stars with Balmer-emission consistent
with magnetically confined circumstellar material.}
{Archival spectroscopic and photometric data were investigated.}
{\object{HR\,7355} is a formerly unknown He-strong star showing Balmer
  emission.  At $V=6.02$\,mag, it is one of the brightest objects
  simultaneously showing anomalous helium absorption and hydrogen
  emission. Among similar objects, only $\sigma$\,Ori~E has so far been
  subjected to any systematic analysis of the circumstellar material
  responsible for the emission.  We argue that the double-wave photometric
  period of 0.52\,d corresponds to the rotation period.  In tandem with the
  high projected equatorial velocity, $v \sin i=320$\,km\,s$^{-1}$, this short
  period suggests that HR\,7355 is the most rapidly rotating He-strong star
  known to date; {a class that was hitherto expected to host stars with
  slow to moderate rotation only.}}
{}
\maketitle
%
\section{Introduction}
In the early B-type spectral range a subclass of He-strong stars is
found, i.e.\ stars showing Helium lines with abnormally large
equivalent widths.  The chemical surface abundances of these stars are
influenced by the presence of a strong magnetic field, resulting in a
He overabundance that typically varies in strength over the stellar
surface.

Because He-strong stars are sufficiently luminous to harbour radiatively
driven winds \citep[as diagnosed by ultraviolet absorption line diagnostics;
see][]{1990ApJ...365..665S}, they represent ideal laboratories for
understanding the process of magnetic wind confinement
\citep[][]{1997A&A...323..121B}.  Typically, the fields of these stars are too
strong for them to be amenable to the magnetohydrodynamical (MHD) simulations
\citep[e,g,][]{2002ApJ...576..413U}. However, an alternative Rigidly Rotating
Magnetosphere (RRM) model for the circumstellar distribution of
magnetocentrifugally confined wind plasma by \citet{2005MNRAS.357..251T} has
shown much promise in reproducing the detailed optical variability of the
archetype emission-line He-strong star \object{$\sigma$\,Ori~E}. 

{To date, our knowledge on He-strong stars is limited to slow to moderate
rotators, as no rapid rotators had been found. This goes as far as to the
conclusion that slow rotation is an intrinsic property of He-strong stars
\citep{1983ApJ...268..195W,1999A&A...345..244Z}, that has to be taken into
account by the search for the origin of the magnetic field.

This work not only reports the discovery of one more bright massive star to
host a magnetosphere for application of the above model, but also extends the
parameter range in which He-strong stars are found by a factor of about two in
rotational velocity space.}


\object{HR\,7355} (\object{HD\,182\,180}, \object{HIP\,95\,408}) is a
little-observed B2Vn star of $6^{\rm th}$ magnitude ($V=6.02, B=5.91$),
located toward the galactic center. It was listed as a MK-standard by
\citet{1969ApJ...157..313H}, but never examined in detail. Other investigators
have instead classified it as B5IV \citep[see][]{1964PLPla..28....1J}. From
studies of larger samples of stars that included HR\,7355, we know that the
star is very rapidly rotating: \citet{2002ApJ...573..359A} measured $v\sin i =
320$\,km\,s$^{-1}$, while \citet{2000AcA....50..509G} report $v\sin i =
270\pm30$\,km\,s$^{-1}$. Hipparcos photometric data indicate that the star is
a periodic variable, with $P=0.26\,$d \citep{2002MNRAS.331...45K}.


\begin{figure*}[t]
%
  \parbox{18cm}{\centerline{\psfig{figure=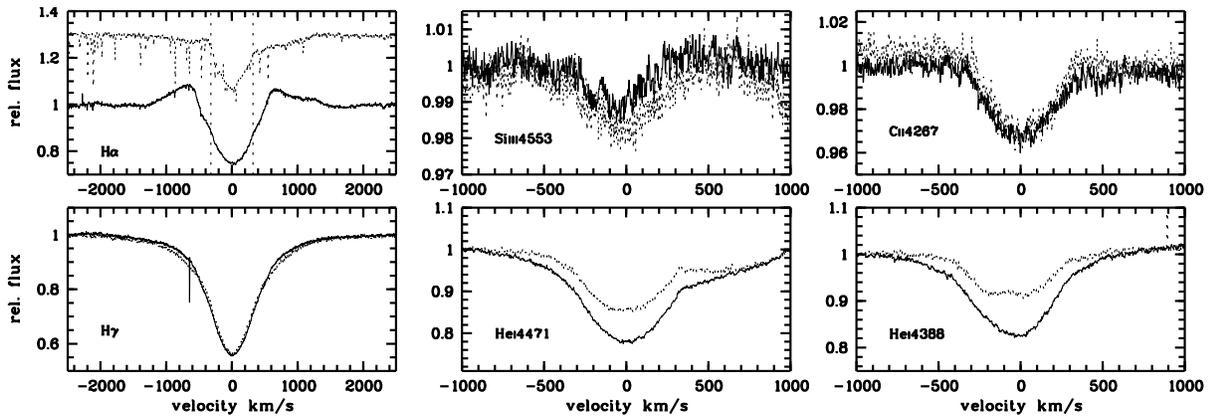,width=16cm,clip=t,angle=0}}}%
\caption[xx]{\label{spectra}Changes in several representative lines
  between the spectra taken in 1999 (solid line) and 2004 (dotted). The 1999
  profile has H$\alpha$ emission extending out to several times beyond $v \sin
  i$ (the latter indicated by the vertical dotted lines, upper left) Note that
  for the Balmer lines (left column) a wider range in velocity is shown than in
  the other panels.}
\end{figure*}

\section{Observations}
During guaranteed-time observations with FEROS at the ESO-1.52m telescope in
1999, that included several magnetic and emission-line stars in the target
list, HR\,7355 was observed once 
on July 25, 1999 (HJD=2451385.507) with $S/N=280$
and noted as a weak emission star, but this result was not published
(Fig.~\ref{spectra}, upper left panel). FEROS is an echelle instrument
covering the spectral range 3600-9200\,\AA\ with a spectral resolving power of
$\Delta\lambda/\lambda=48\,000 $ \citep{1999Msngr..95....8K}. When re-scanning
available FEROS spectra of Be stars in a search for candidate hot stars having
magnetospheres, we noticed that a second FEROS spectrum of HR\,7355 had been
obtained on July 05, 2004 (HJD=2453191.879) with $S/N= 270$
\citep{2005A&A...437..467E}.
The latter spectrum has been retrieved from the ESO-science archive and
reduced with the FEROS standard Data Reduction System, available from ESO.
The two spectra suffer from a somewhat imperfect continuum
normalization, not untypical for echelle data. This tends to limit the
accuracy of equivalent width measurements for Stark broadened lines such as
the Balmer lines, and for hot stars such as HR\,7355, Helium lines can
similarly be affected.


\section{Analysis}

\subsection{Photometry}

Photometric data have been obtained from the Hipparcos satellite
archive. Two points were removed as outliers, leaving 41 remaining
photometric measurements spanning the interval from JD=2447967 to
JD=2449061.

We repeated the analysis by \citet{2002MNRAS.331...45K} and were able
to confirm their period for sinusoidal variations: $P_{\rm
sin}=0.260714\pm0.000003$\,d. Subtracting this peak from the Fourier
spectrum significantly decreases the overall power in the variability,
and the remaining, second-strongest peak is close to the first harmonic
of the strongest one.

\subsection{Spectroscopy}

Figure \ref{spectra} shows a selection of the spectral
lines observed.  Most intriguing are the changes in the He{\sc i}
lines, and the width of the Balmer emission.

In classical Be stars \citep{2003PASP..115.1153P}, where the emission
arises from a Keplerian disk having near-circular particle orbits, the
highest kinematic velocity possible is the orbital velocity at the
stellar surface, typically from a few to about five hundred
km\,s$^{-1}$ for late and early type B-stars, respectively. In the
present case, however, the Balmer emission extends from $-1350$ to
$+1500$\,km\,s$^{-1}$. For a strong emission line, peaking at several
times the flux of the local continuum, scattering processes can
broaden its base; however, the line seen in HR\,7355 is certainly too
weak for such broadening to be occurring. We are thus led to to
conclude that the extent of the Balmer emission in HR\,7355 is
governed by the non-Keplerian kinematics of the emitting material
itself.  Between 1999 and 2004, the emission decreased {strongly}
both in strength and kinematic width, {but remained present as
  distinct peak on the blue side slightly outside $v \sin i$, yet less
  obvious as a filling-in of the absorption on the red side.}

The photospheric absorption lines also differ between the two epochs.
The He{\sc i} lines show the most striking variations -- the
equivalent widths (EWs) of some change by more than a factor of 2 (see
Table~\ref{abundances}). The changes arise across the entire line
width, affecting the Stark broadening wings as well as the line
cores. Such behaviour cannot be explained by pulsation; radial
pulsation displaces the entire line, while non-radial pulsation
distorts the profile within the limits of $v \sin i$, but tends to
conserve the total EW. Neither behaviour is consistent with that seen
in HR\,7355. To the contrary, the He{\sc i} profiles obtained in 2004,
in particular the 4388 and 4713 lines (Fig.~\ref{spectra}), resemble
the signature of a spot on a rotating star (cf. Sect.~\ref{similarity}).

Variations are also seen in lines other than He{\sc i}, but they are
much weaker. The wings of the Balmer lines are somewhat deeper in the
spectrum taken 2004 than in the one from 1999, and the EW of Si{\sc
  iii} is slightly larger, {i.e.\ by about $1.6\sigma$, see
  Table~\ref{abundances}. The uncertainty of the latter measurement
  has been estimated from a series of manual and semi-automated
  measurements, with integration limits chosen to embrace all
  potential continuum normalization errors, so $\sigma=25\,$m\AA\ is
  probably conservative.}  There are also some lines that do not
change: The C{\sc ii}\,4267 profile remains unchanged within the
limits given by noise and normalization, and the apparent variations
of the Mg{\sc ii}\,4481 profile can be entirely attributed to the
changes in the neighbouring He{\sc i}\,4471 line.

\section{Discussion}

\subsection{The rotational period}

\begin{figure}[t]
  \parbox{8.8cm}{\centerline{\psfig{figure=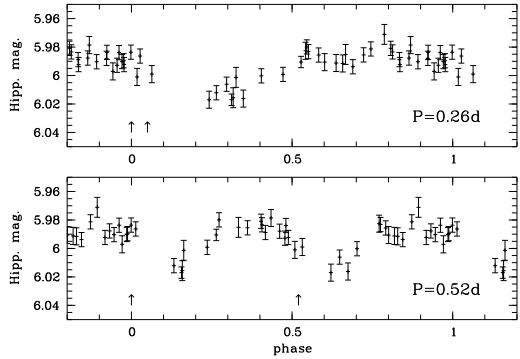,width=7.5cm,clip=t,angle=270}}}
\caption[xx]{\label{hipphot}The Hipparcos photometry sorted with the
  single and double-wave periods. As epoch the date of the 1999 FEROS
  spectrum has been chosen; the respective phases of both spectra are
  indicated by arrows.}
\end{figure}

The high projected equatorial velocity of HR\,7355\ suggests that we
see the star close to equator-on. If an oblique-dipole magnetic field
is responsible for the confinement of the Balmer-emitting
circumstellar material, the density of this material should be highest
at the twin intersections between the magnetic and rotational equators
(see Sect.~\ref{magnetosphere}, {also for an explanation of the
  H$\alpha$ variations, which are rather long-term changes rather than
  being rotational}). When viewed from an equator-on aspect,
these high-density regions will transit the stellar disk twice per
rotation cycle, leading to a double-wave photometric signature. In
this scenario, the rotation period is identified as twice the observed
sinusoidal value, i.e.\ $P_{\rm dw}=0.521428\pm0.000006$\,d.  The
Hipparcos photometry phased with both of those periods is shown in
Fig.~\ref{hipphot}, with the date of the 1999 spectrum,
HJD=2\,451\,385.507, being adopted as phase zero.

Even with the large temporal separation of the spectra, of almost 5
years, the accuracy of the period above is sufficient to phase the
spectra with acceptable uncertainty. Adopting the sinusoidal period,
the two spectra are $6929.05\pm0.07$ cycles apart, i.e.\ having almost
the same phase. If we are to seek a common origin for the
spectroscopic and photometric variations, this small phase difference
is not compatible with the strong changes seen in the
spectra. However, with the double wave period, the cycle separation
becomes $3464.52\pm0.04$ -- a half-phase difference, which is much
more plausible.

The two IUE spectra of HR\,7355, SWP39549 and 39556, do not show any
significant differences. The spectra are separated by 8.879,\,d; with
the above periods this corresponds either to 34.06 or to 17.03 cycles,
both values having a small phase difference consistent with the
absence of variations.

\subsection{Similarity to $\sigma$\,Ori~E} \label{similarity}

The changes in the He{\sc i} lines seen in Fig.~\ref{spectra} are a
clear indicator of abundance variations across the surface. In
particular, the extent of the variability across the entire width of
the lines, including the Stark-broadened wings, can hardly be
attributed to any other mechanism. The same kind of abundance
variations can be seen in \object{$\sigma$\,Ori~E}; there, the He{\sc i}
equivalent width changes in anti-phase compared to lines of Carbon,
Oxygen, Silicon, and Magnesium. The Hydrogen lines of $\sigma$\,Ori~E
are also modulated, however in a more complicated fashion due to a
combination of photospheric and circumstellar effects. In general, the
variations of HR\,7355, inasmuch that they can be estimated from only
two spectra, are quite similar to those seen in $\sigma$\,Ori~E, as
reported e.g.\ by \citet{2000A&A...363..585R}.

\begin{table}[t]
  \caption[xx]{\label{abundances}Equivalent widths of
    selected lines, compared to equivalent widths measured
    from model spectra by \citet[][see Sect.\
    \ref{physprop} for details]{2000A&A...363.1051Z}. 
Measurements are given in m\AA, and the
    typical error in the observed values are about 10\,\% for the strong 
lines of H and He, and about 25\,m\AA\ for the weak lines of C{\sc ii} 
and Si{\sc iii}.}
\begin{center}
\begin{tabular}{l|rrrr}
  Line             & 1999 & 2004 & He/H=0.4 & He/H=0.1\\
  \hline
  H$\gamma$        & 6300 & 6600 & 6640 & 6460 \\
  He{\sc i}\,4713  & 430  & 230  & 470  & 306 \\
  He{\sc i}\,4388  & 1650 & 760  & 1590 & 840 \\
  He{\sc i}\,5015  & 490  & 290  & 494  & 318 \\
  {C{\sc ii}\,4267}  & 187  & 188  & --  & -- \\
  {Si{\sc iii}\,4553}& 85  & 124  & --  & -- \\
\end{tabular}
\end{center}
\end{table}

\subsection{\label{physprop}Physical properties of HR\,7355}

Due to the significant effects of rotational broadening, only the
strongest photospheric lines can easily be measured and used to
constrain the star's fundamental parameters. Unfortunately, second
only to the (emission-contaminated) Balmer lines in this respect are
the He{\sc i} lines, and these of course are strongly
variable. Nevertheless, at least rough estimates can be attempted.

The luminosity classification as a dwarf star and the broadening wings
of the Hydrogen lines are consistent with a surface gravity $\log
g\approx4.0$. Using this as a reference point, the spectral features,
including the equivalent widths of the Balmer lines, point toward an
effective temperature on the order of 20\,000\,K.  This value is
bracketed by the published spectral classifications, B2V
\citep{1969ApJ...157..313H} and B5IV \citep{1964PLPla..28....1J}, but
is closer to B2V {\citep[see Table 10 of][for instance]{2007A&A...471..625T}}.

The very high projected equatorial velocity of $v \sin
i=320$\,km\,s$^{-1}$ makes HR\,7355 the most rapidly rotating
He-strong star known to date.  Statistically, the class of He-strong
stars is deficient in rapid rotators
\citep{1983ApJ...268..195W,1999A&A...345..244Z}, with $\sigma$\,Ori~E
($v \sin i=165$\,km\,s$^{-1}$ {and $i\approx75^\circ$} ) being one of
the record holders so far.  Assuming $\sin i=1$ and identifying the
double wave period as rotational, the stellar radius of HR\,7355
{would be about 3.3\,R$_\odot$, too small for a B2V star even at the
  ZAMS \citep{1995MNRAS.277.1547B}}.  However, HR\,7355 is rotating
sufficiently close to the critical limit that gravity darkening could
bias the measurement of $v \sin i$ toward lower apparent values
\citep{2004MNRAS.350..189T}.

To derive a preliminary estimate of the range in surface Helium
abundances, we compared the FEROS spectra against models published by
\citet{2000A&A...363.1051Z}. These models use He/H abundance ratios
ranging from 0.1 to 1.0 for a sequence of effective temperatures and
surface gravities, including $T_{\rm eff}=20\,000$\,K, $\log
g=4.0$. The results from equivalent width measurements for a few lines
in the observed spectra are compared in Table~\ref{abundances} against
values measured from the models for the above parameters and He/H=0.1
and 0.4.  The typical EW-error of 10\,\% was derived by a conservative
estimate from repeated manual measurements.
%
%
We conclude that the disk-averaged stellar hemisphere observed in 1999
was enriched in Helium by a factor of about 4, while the one observed
in 2004 was about normal or even slightly depleted in Helium. A more
detailed study is needed to refine these numbers, however.

\subsection{The putative magnetosphere of HR\,7355} \label{magnetosphere}

The absorption-line changes of HR\,7355 are a clear indication of
spatial structure in the surface abundances at the very least of
Helium, Silicon, and possibly Hydrogen. For B-type stars, this
structure is the typical signature of a strong magnetic field, on the
order of several kiloGauss. The presence of emission -- with an
extension out to almost $\pm$1500\,km\,s$^{-1}$ that is more likely to
be kinematic rather than due to scattering -- lends independent
support to the presence of a strong field that is able to confine
circumstellar plasma and torque it into co-rotation.

The RRM model \citep{2005MNRAS.357..251T} assumes a magnetic field
sufficiently strong that the circumstellar environment is completely
dominated by the field, i.e.\ wind plasma upflowing from the star is
forced to follow the field lines, but does not influence them. The
model predicts the steady accumulation of plasma at points along field
lines where the effective (gravitational plus centrifugal) potential
is at a local minimum. For an oblique dipole field -- the sort most
commonly detected in chemically peculiar stars -- the locus formed by
such minima resembles a warped disk; moreover, the distribution of
accumulated plasma in this disk is concentrated into two elongated
cloud-like regions, centered along the twin intersections between
magnetic and rotational equators.

The RRM model predicts a distinctive observational signature for the
warped magnetospheric disk. Because it co-rotates with the star, it
exhibits double-peaked emission with a strength that varies due to
optical depth and occultation effects. Depending on rotational and
magnetic inclination, the disk may transit the stellar disk either
once or twice per rotation cycle, in both cases absorbing photospheric
flux. In HR\,7355, the equator-on aspect means that two such eclipses
should occur per rotation cycle \citep[see][]{2007Townsendb}, in
accordance with our assumption that the double-wave period corresponds
to the rotation period.

The changes in the total emission strength between 1999 and 2004
cannot be constrained to a short or long timescale from the two
observations alone. However, they are not easily ascribed to any
short-periodic type of variation, and are more likely to have occured
on longer timescales, for instance during a breakout of accumulated
material from its magnetic confinement (see Appendix of
\citealp{2005MNRAS.357..251T}; see also
\citealp{2006ApJ...640L.191U}).

\subsection{Other He-strong emission line stars}

In the handful of other He-strong stars showing convincing cases of
H$\alpha$ emission -- \object{$\delta$\,Ori~C}, \object{V\,1046 Ori}
and \object{HD\,64\,740} -- there are only brief reports noting the
emission and its variability, but no in-depth investigations have been
published so far
\citep{1974ApJ...191L..95W,1979A&AS...35..313P,1991JRASC..85..202B,1998A&A...337..183B}. Further
spectra are required for those to obtain a sufficient database, both
in amount and quality, for a detailed study with current models.  A
few more candidates mentioned by \citet{1997A&A...324..949Z} are
suspected candidates on the basis of photometry alone or might show
nebular H$\beta$ emission instead of circumstellar one.

\section{Conclusions}

HR\,7355 is a previously unknown spectroscopically variable star, and
as such it should no longer be used as a spectral standard star. In
its capacity as the newest member of the He-strong class, it is not
only one of the brighter stars in this class, but is also the most
rapidly rotating.

In addition to its spectral variability, HR\,7355 is periodically variable in
photometry, with either a single-wave sinusoidal lightcurve of $P_{\rm
sin}=0.260714\pm0.000003$\,d or a double-wave pattern with $P_{\rm
dw}=0.521428\pm0.000006$\,d.  At this point we cannot firmly exclude the
possibility that the photometric variations originate in some mechanism other
than the surface abundance inhomogeneities. However, the spectra do not show
the typical signature of pulsation, and moreover we do not find any other
signal in the photometric data consistent with the rotation period. Thus, if
the spectral and photometric variations repeat on the same period, then that
period is the double-wave period, which also is the rotational period. {It
is the first rapidly rotating He-strong star, and as such may pose a challenge
to field origin hypotheses that would have led to strong magnetic braking.}

We intend to begin a monitoring campaign on the star, to obtain a
spectroscopic time series {for further analysis of the fundamental
parameters of HR\,7355, as well as for application of} the framework of the
RRM model of \citet{2005MNRAS.357..251T}. This model has proven extremely
successful in explaining the emission-line variations of $\sigma$\,Ori~E
\citep{2005ApJ...630L..81T}, and we are optimistic that it can explain the
behaviour of HR\,7355 too.

\begin{acknowledgements}
  RHDT is supported by NASA grant LTSA/NNG05GC36G.  We thank the FEROS
  consortium for observing HR\,7355 in 1999, during the time
  guaranteed to the consortium for building the instrument.  This
  study made use of the SIMBAD and ADS databases, as well of the ViZiR
  catalog services. We thank the second referee, C.~J.~Evans, for his
  suggestions how to improve the presentation of this work.
\end{acknowledgements}

\bibliographystyle{aa}
\bibliography{7355brich}

\clearpage

\end{document}